\documentclass[aps,prl,twocolumn,showpacs]{revtex4}
\usepackage{graphicx}
\usepackage{dcolumn}
\usepackage{bm}
\newcommand{\bk}{{\bf k}}
\newcommand{\pd}{$P^{ave}_d$ }

\newcommand{\M}{$\langle M\rangle$ }

\begin{document}

\title{Absence of Coexistence of Superconductivity and Antiferromagnetism of the Hole-Doped Two-Dimensional Extended $t-J$ Model}

\author{C. T. Shih}
\affiliation{Department of Physics, Tunghai University, Taichung,
Taiwan}

\author{Y. C. Chen}
\affiliation{Department of Physics, Tunghai University, Taichung,
Taiwan}

\author{C. P. Chou}
\affiliation{Department of Physics, National Tsinghua University,
Hsinchu, Taiwan}

\author{T. K. Lee}
\affiliation{Institute of Physics, Academia Sinica, Nankang,
Taiwan}

\date{\today}

\begin{abstract}
The possibility of coexistence of superconductivity (SC) and
antiferromagnetic long range order (AFLRO) of the two-dimensional
extended $t-J$ model in the very underdoped region is studied by
the variational Monte-Carlo (VMC) method. In addition to using
previously studied wave functions, a recently proposed new wave
function generated from the half-filled Mott insulator is used.
For hole-doped systems, the phase boundary between AFLRO and
$d-$wave SC for the physical parameters, $J/t=0.3$, $t'/t=-0.3$
and $t''/t=0.2$, is located near hole density $\delta_c = 0.06$,
and there is {\it no} coexistence. The phase transition is
first-order between these two homogeneous phases at $\delta_c$.
\end{abstract}


\maketitle

Correlation between the $d$-wave SC and AFLRO is one of the
critical issues in the physics of the high-temperature
superconductivity (HTS)\cite{anderson97,zhang97}. Early
experimental results showed one of the common features of the HTS
cuprates is the existence of AFLRO at temperature lower than the
N$\acute{e}$el temperature $T_N$ in the insulating perovskite
parent compounds. When charge carriers (electrons or holes) are
doped into the parent compounds, AFLRO is destroyed quickly and
then SC appears. In most thermodynamic measurements, AFLRO does
not coexist with SC\cite{kastner98}. However, this is still a
controversial issue. Recent experiments such as neutron-scattering
and muon spin rotation show that the spin density wave (SDW) may
compete, or coexist with SC under the external magnetic field
\cite{lake02,miller02,sonier03}. Remarkably, elastic neutron
scattering experiments for underdoped $YBa_2Cu_3O_x$ ($x=6.5$ and
$6.6$, $T_c=55K$ and $62.7K$, respectively) show that the
commensurate AFLRO develops around room temperature with a large
correlation length $\sim 100\AA$ and a small staggered
magnetization $m_0\sim 0.05\mu_B$\cite{sidis01,hodges02,mook02}.
These results suggest that AFLRO may coexist with SC but the
possibility of inhomogeneous phases is not completely ruled out.

For the theoretical part, the two-dimensional (2D) t-J model is the
first model proposed\cite{anderson87} to understand the physics of
HTS. Anderson proposed
the resonating-valence-bond (RVB) theory for the model about
one and a half decades ago. The theory is reexamined
again\cite{anderson03} recently. The authors compared the prediction of the
RVB theory with several experimental results and found the
theory to have successfully explained the main features of cuprates.
This so called ``plain vanilla'' theory did not consider the issue
of AFLRO, which must be addressed at very low
doping.
From analytical and numerical studies of the $t-J$ model, it was
shown that at half-filling, the $d-$wave RVB state with AFLRO is a
good trial wave function (TWF). In this case, SC correlation is zero because
of the constraint of no-double-occupancy. Upon doping, the
carriers become mobile and SC revives while AFLRO is quickly
suppressed. However, if the doping density is still small, AFLRO
will survive. Thus SC and AFLRO coexist in the very underdoped
regime\cite{lee88,chen90,giamarchi91,inaba93,himeda99}.
Exact diagonalization (ED) up to 26
sites show that both SC and AFLRO are enhanced by the external
staggered field. This result also implies these two orders can coexist
homogeneously in a 2D $t-J$ model\cite{saiga03}.
However, the regime
of AFLRO predicted by these studies extend to larger doping than
the experimental results. The robustness of the coexistence of SC
and AFLRO seems to be inconsistent with experiments\cite{varma03}.

There are several experimental and theoretical studies suggesting
the presence of the next- and third-nearest-neighbor hopping terms
$t'$ and $t''$ in cuprates. For example, the topology of the large
Fermi surface and the single-hole dispersion studied by ARPES
\cite{damascelli03}, and the asymmetry of phase diagrams of the
electron- and hole-doped cuprates can be understood by introducing
these terms. Further, these longer range hopping terms may be
essential for the large enough $T_c$ for the $t-J-$type
models\cite{shih98,shih04}. In this paper, we'd like to
demonstrate the phase diagram constructed by VMC results of the
extended $t-J$ model. The trial WF's for very underdoped systems
are generalized from the single-hole and slightly-doped WF
proposed by Lee {\it et al.}\cite{lee97,lee03}. The results for
the hole-doped case show there is {\it no} coexistence of $d-$wave
pairing and AFLRO when the next- and third-nearest neighbor
hopping terms are introduced. And the phase boundary of AFLRO is
pushed to lower doping density.

The Hamiltonian of the extended $t-J$ model is
\begin{eqnarray}
&H&=H_t+H_J=\\
&-&\sum_{ij}t_{ij}(\tilde{c}^\dagger_{i,\sigma}\tilde{c}_{j,\sigma}
+ H.C.)+J\sum_{<i,j>}({\bf{S_i}\cdot
S_j}+\frac{1}{4}n_in_j)\nonumber \label{e:tjm}
\end{eqnarray}
where $t_{ij}=t$, $t'$, and $t''$ for sites $i$ and $j$ are
nearest, next nearest, and the third nearest neighbors. $<i,j>$ in
$H_J$ means the spin-spin interaction occurs only for nearest
neighbors. $\tilde{c}_{i,\sigma}=(1-n_{i,-\sigma})c_{i,\sigma}$,
satisfies the no-double-occupancy constraint. At half-filling, the
system is reduced to the Heisenberg Hamiltonian $H_J$. As carriers
are doped into the parent compound, $H_t$ is included in the
Hamiltonian.

First we exam the phase diagram of the $t-J$ model, that is,
$t'=t''=0$. Following Ref.\cite{inaba93,lee97}, three mean-field
order parameters are introduced: the staggered magnetization $m_s
= \langle S_A^z\rangle = -\langle S_B^z\rangle$, where the lattice
is divided into A and B sublattices, the uniform bond order
parameters $\chi=\langle \sum_\sigma
c^\dagger_{i\sigma}c_{j\sigma}\rangle$, and $d-$wave RVB ($d-$RVB)
one $\Delta=\langle
c_{j\downarrow}c_{i\uparrow}-c_{j\uparrow}c_{i\downarrow}\rangle$
if $i$ and $j$ are n.n. sites in the $x$ direction and $-\Delta$
for the $y$ direction. The Lee-Shih WF, which is the mean-field
ground state WF is
\begin{equation}
\mid \Psi_{LS}\rangle = P_d\left(\sum_{\bk\in SBZ}(A_\bk
a^\dagger_{\bk\uparrow}a^\dagger_{-\bk\downarrow}+ B_\bk
b^\dagger_{\bk\uparrow}b^\dagger_{-\bk\downarrow})\right)^{N_s/2}\mid
0\rangle \label{e:ls_twf}
\end{equation}
where $N_s$ is the total number of sites and
$A_\bk=(E_\bk^{(1)}+\xi_\bk^-)/\Delta_\bk$ and
$B_\bk=-(E_\bk^{(2)}-\xi_\bk^+)/\Delta_\bk$ with
$E_\bk^{(1)}=(\xi^-_\bk{^2} + \Delta^2_\bk)^{1/2}$ and
$E_\bk^{(2)}=(\xi^+_\bk{^2} + \Delta^2_\bk)^{1/2}$. Here
$\Delta_\bk=\frac{3}{4}\Delta (cosk_x-cosk_y)$. Energy dispersions
for the two SDW bands are $\xi_\bk^\pm =
\pm[(\epsilon_\bk+\mu)^2+(Jm_s)^2]^{1/2}-\mu$ with $\epsilon_\bk =
-2(t\delta+\frac{3}{8}J\chi)(cosk_x+cosk_y)$.
$a_{\bk\sigma}=\alpha_\bk c_{\bk\sigma} + \sigma\beta_\bk
c_{\bk+{\bf Q}\sigma}$ and $b_{\bk \sigma} = -\sigma\beta_\bk
c_{\bk\sigma}+\alpha_\bk c_{\bk+{\bf Q}\sigma}$, where ${\bf
Q}=(\pi, \pi)$, $\alpha^2_\bk =
\frac12\{1-[(\epsilon_\bk+\mu)/(\xi^+_\bk+\mu)]\}$ and
$\beta^2_\bk=\frac12\{1+[(\epsilon_\bk+\mu)/(\xi^+_\bk+\mu)]\}$,
are the operators of the lower and upper SDW bands, respectively.
$\mu$ is the chemical potential which determines the number of
electrons. Note that the summation in Eq.(\ref{e:ls_twf}) is taken
over the sublattice Brillouin zone (SBZ). The operator $P_d$
enforces the constraint of no doubly occupied sites for cases with
finite doping.

For the half filled case, $\mu=0$ and the optimal variational
energy of this TWF obtained by tuning $\Delta$ and $m_s$ in the
VMC simulation is $-0.332J$ per bond which is within $1\%$ of the
best estimate of the ground state energy of the Heisenberg model.
For the case of pure AFLRO without $\Delta$, energy per bond is
about $3$ to $4\%$ higher. Upon doping, there are two methods to
modify the TWF: one is to use the SDW bands with a nonzero $\mu$,
the other is to create charge excitations from the half-filled
ground states. For the former method, the TWF is optimized by
tuning $\Delta$, $m_s$ and $\mu$. Note that for larger doping
densities, AFLRO disappears ($m_s=0$) and the WF reduces to the
standard $d-RVB$ WF. For the latter method, the WF is the ``small
Fermi pocket'' state $\mid \Psi_p\rangle$:
\begin{equation}
\mid \Psi_{p}\rangle = P_d\left(\sum_{\bk\in SBZ, \bk\notin Q_p
}(A_\bk a^\dagger_{\bk\uparrow}a^\dagger_{-\bk\downarrow}+ B_\bk
b^\dagger_{\bk\uparrow}b^\dagger_{-\bk\downarrow})\right)^{N_s/2}\mid
0\rangle \label{e:pocket}
\end{equation}
$\bk\notin Q_p$ means the $\bk$ points in the Fermi pocket $Q_p$
are not occupied. For example, for 4 holes in $12\times 12$
lattice, $Q_p=\{(\pi/2,\pi/2),(\pi/2,-\pi/2)\}$. The number of
holes is twice of the number of \bk-points in $Q_p$ and $\mu$ is
identical to zero in Eq.(\ref{e:pocket}). In general, for the ground state
the set
$Q_p$ should be determined variationally. Yet as we expect, it
agrees well with the rigid band picture in the slightly doped
cases as in Ref.\cite{lee03}.

\begin{figure}[here]
\rotatebox{-90}{\includegraphics[width=2.2in]{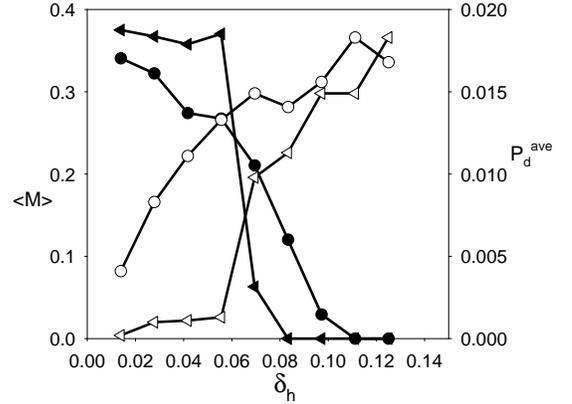}}
\caption{\M (full symbols) and \pd (empty symbols) for $J/t=0.3$,
$t'=t''=0$ (circles) and $t'/t=-0.3$, $t''/t=0.2$ (triangles) for
hole doped $12\times 12$ lattice.} \label{f:phase_144}
\end{figure}

The staggered magnetization $\langle
M\rangle=\frac{1}{N_s}\langle\sum_je^{i{\bf Q}\cdot{\bf
R_j}}S^z_{\bf R_j}\rangle$ and the d-wave pair-pair correlation
$P_d({\bf R}) =\frac{1}{N_s}\langle\sum_i\Delta^\dagger_{\bf
R_i}\Delta_{\bf R_i+R}\rangle$, where $\Delta_{\bf R_i}=c_{{\bf
R_i}\uparrow}(c_{{\bf R_i+\hat{x}}\downarrow}+c_{{\bf
R_i-\hat{x}}\downarrow}-c_{{\bf R_i+\hat{y}}\downarrow}-c_{{\bf
R_i-\hat{y}}\downarrow})$ are measured for $J/t=0.3$ and
$t'=t''=0$ for the $12\times 12$ lattice with periodic boundary
condition. $P^{ave}_d$ is the averaged value of the long-range
part ($\mid {\bf R}\mid > 2$) of $P_d({\bf R})$. The resulting \M
(full circles) and \pd (empty circles) are shown in
Fig.\ref{f:phase_144}.

It can be seen in Fig.\ref{f:phase_144} that in the underdoped
region for the $J/t=0.3$, $t'=t''=0$ case, AFLRO (full circles)
coexists with SC (open circles) for $\delta_c\le 10\%$, which is
smaller than the weak-coupling mean-field result $\sim
15\%$\cite{inaba93}, but still larger than the phase boundary of
AFLRO determined by experiments $\delta_c < 5\%$. The energies of
$\mid\Psi_{LS}\rangle$ are lower than those of
$\mid\Psi_{P}\rangle$ for all doping densities in this case. This result is also
consistent with the results reported by Himeda and Ogata\cite{himeda99}.
The VMC result is more realistic than the weak-coupling one. It may
result from the rigorous no-double-occupancy constraint that
suppresses the AFLRO faster than the constraint-relaxed mean-field
approximation.

Now we examine the phase diagram for $J/t=0.3$, $t'/t=-0.3$ and
$t''/t=0.2$. For this case, the WF Eq.(\ref{e:ls_twf}) is modified
by replacing $\mu$ by
$\mu+4t'_vcosk_xcosk_y+2t''_v(cos2k_x+cos2k_y)$ due to the second
and third nearest neighbor hopping terms. $t'_v$ and $t''_v$ are
variational parameters. $t'_v$ and $t''_v$ are not necessarily
equal to the bare values $t'$ and $t''$ because the constraint
strongly renormalizes the hopping amplitude. On the other hand,
the effect of $t'$ and $t''$ on $\mid\Psi_{P}\rangle$ is the
choice of \bk-points in $Q_p$, and the form of Eq.(\ref{e:pocket})
is not changed.

\begin{figure}[here]
\rotatebox{-90}{\includegraphics[width=2.2in]{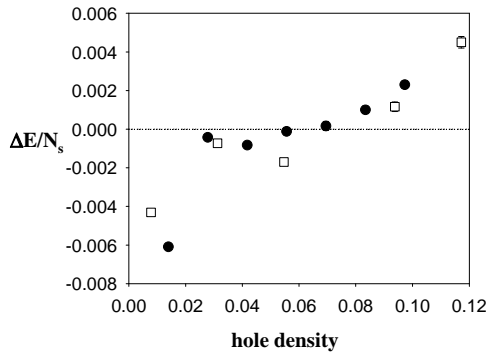}}
\caption{Energy difference per site between the two wave functions
$\mid \Psi_{P}\rangle$ and $\mid\Psi_{LS}\rangle$ for $t'/t=-0.3$
and $t''/t=0.2$ in $12\times 12$ (full circles) and $16\times 16$
(open squares) lattices.}\label{f:eng_diff}
\end{figure}

The optimal wave functions for different densities are determined
by minimizing the variational energies among $\mid\Psi_p(m_s,
\Delta, \{Q_p\})\rangle$ and $\mid
\Psi_{LS}(m_s,\Delta,t'_v,t''_v,\mu)\rangle$. The differences of
the energies of best $\mid\Psi_{P}\rangle$ and
$\mid\Psi_{LS}\rangle$ for various hole densities are shown in
Fig.\ref{f:eng_diff}. \M (full triangles) and \pd (open triangles)
for $12\times 12$ lattice are shown in Fig.\ref{f:phase_144}.

\begin{figure}[here]
\rotatebox{-90}{\includegraphics[width=2.2in]{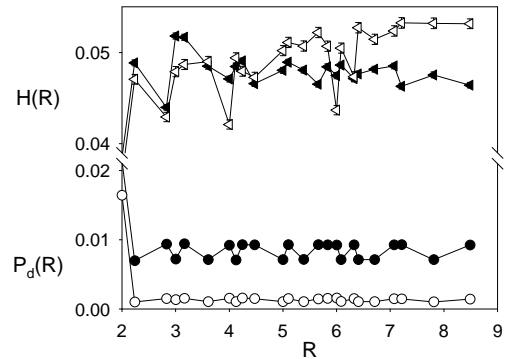}}
\caption{Pair-pair correlation $P_d(R)$ (circles) and hole-hole
correlation $H(R)$ (triangles) of $\mid \Psi_{LS}\rangle$ (full
symbols) and $\mid \Psi_p\rangle$ (open symbols) for 8 holes in a
$12\times 12$ lattice.} \label{f:corr}
\end{figure}

It can be seen that level crossing occurs at $\delta_h \sim 0.06$.
$\mid\Psi_{P}\rangle$ has lower energy below the critical density.
To show $\mid\Psi_{LS}\rangle$ and $\mid\Psi_p\rangle$ belong to
two different types of WF, we calculate the overlap of them.
($\frac{\langle \Psi_{LS}\mid \Psi_p\rangle}{\mid \Psi_{LS}
\mid\mid \Psi_{p}\mid }$) is only $0.0113(4)$. The almost
orthogonality of the two wave functions implies that the ground
state WF's switch at the critical density.
Another evidence is shown by the correlation functions of the two wave
functions shown in Fig.\ref{f:corr}. It is clear that the holes in
$\mid \Psi_p\rangle$ repel each others and pairing is very small,
while the behavior is opposite for $\mid \Psi_{LS} \rangle$.

For $\delta_h < 0.06$, $\mid\Psi_p\rangle$ is the ground state WF
and \M is a little larger than the $t'=t''=0$ case while \pd is
suppressed by one order of magnitude. Thus there is AFLRO but {\it
no} SC in this regime. The behavior is quite different from
$\mid\Psi_{LS}\rangle$ for the same doping regime for $t'=t''=0$
case, whose \pd coexists with \M. The possible reason is that the
WF $\mid\Psi_p\rangle$ gains energy (short-range effect) from its
$d-RVB$ feature as $\mid\Psi_{LS}\rangle$, yet $P^{ave}_d$ is
greatly suppressed by replacing $\mu$ by $Q_p$ to control the
density. This replacement seems to make the WF decoherent for pairing.

For $\delta_h$ larger than $0.06$, the RVB state ($m_s=0$ in $\mid
\Psi_{LS}\rangle$) optimizes the energy. \pd increases and \M
drops to zero sharply. Unlike the $t'=t''=0$ case there is no
region optimized by $\mid \Psi_{LS}\rangle$ with non-zero $m_s$.
In conclusion, there is no coexistence of AFLRO and SC for the
$t'/t=-0.3, t''/t=0.2$ case. These parameters are close to the
values for YBCO and BSCO compounds\cite{pavarini}.

In summary, for the extended $t-J$ model, we proposed a new WF
$\mid\Psi_p\rangle$ for the underdoped regime which has lower
variational energy than the traditional WF with coexisting AFLRO
and SC. This WF is constructed under the framework of RVB. The new
wave function has AFLRO but SC is largely suppressed and there is
no coexistence of AFLRO and SC in the underdoped regime of the
hole-doped extended $t-J$ model. The variational phase diagram
shows better agreement with experimental results for the
underdoped HTS cuprates.

Note that in this study we only consider the homogeneous states.
Since the phase
transition comes from the level crossing of the two classes of
states at the critical density $\delta_c=0.06$, it is a first
order phase transition. It is quite natural to have inhomogeneity in the system
near the critical point\cite{burgy01}. It may also lead to other
more novel inhomogeneous states such as stripe phase\cite{himeda02}.
Another interesting result of our study is that the non-coexistence
of SC and AFLRO is much more robust for systems with larger
values of $t'/t$ and $t''/t$ such as YBCO and BSCO\cite{pavarini}. For LSCO where
$t'/t$ and $t''/t$ are smaller, the tendency toward coexistence is larger and the
possibility of inhomogeneous phase will become much more likely.


The work is supported by the National Science Council in Taiwan
with Grant no. NSC-92-2112-M-029-010-, 92-2112-M-029-005-, and
92-2112-M-011-005. Part of the calculations are performed in the
IBM P690 in the Nation Center for High-performance Computing in
Taiwan, and the PC clusters of the Department of Physics and
Department of Computer Science and Engineering of Tunghai
University, Taiwan. We are grateful for their help.

\end{document}